\documentclass{webofc}

\usepackage[varg]{txfonts}   %
\usepackage{cleveref}
\usepackage{hyperref}
\usepackage{url}
\hypersetup{colorlinks=true,citecolor=blue,urlcolor=blue,linkcolor=blue}

\crefname{figure}{Fig.}{Figs.}
\Crefname{figure}{Figure}{Figures}
\crefname{equation}{Eq.}{Eqs.}
\Crefname{equation}{Equation}{Equations}
\crefname{section}{Sec.}{Secs.}
\Crefname{section}{Section}{Sections}
\creflabelformat{equation}{#2\textup{#1}#3}

\makeatletter

\newcommand{\coll}[2]{%
	$\coll@process{#1} + \coll@process{#2}$%
}

\newcommand{\collThree}[3]{%
	$\coll@process{#1} / \coll@process{#2} + \coll@process{#3}$%
}

\newcommand{\collFour}[4]{%
  $\coll@process{#1} / \coll@process{#2} / \coll@process{#3} + \coll@process{#4}$%
}

\newcommand{\coll@process}[1]{%
  \ifx#1A%
    #1%
  \else\ifx#1B%
    #1%
  \else\ifx#1x%
    #1%
  \else\ifx#1p%
    #1%
  \else\ifx#1d%
    #1%
  \else
    \coll@checkHeThree{#1}%
  \fi\fi\fi\fi\fi%
}

\newcommand{\coll@checkHeThree}[1]{%
  \ifnum\pdfstrcmp{#1}{He3}=0 %
    {}^3\mathrm{He}%
  \else
    \mathrm{#1}%
  \fi
}

\makeatother

\begin{document}
\bibliographystyle{apsrev-4-2} %
\title{Heavy Flavour Energy Loss in Small and Large Systems}

\author{\firstname{Coleridge} \lastname{Faraday}\inst{1}\fnsep\thanks{\email{frdcol002@myuct.ac.za}} \and
        \firstname{W.\ A.} \lastname{Horowitz}\inst{1,2}\fnsep
}

\institute{Department of Physics\char`,{} University of Cape Town\char`,{} Private Bag X3\char`,{} Rondebosch 7701\char`,{} South Africa
\and
Department of Physics\char`,{} New Mexico State University\char`,{} Las Cruces\char`,{} New Mexico\char`,{} 88003\char`,{} USA
          }

\abstract{
	We present suppression results for high-$p_T$ $D$ and $\pi$ mesons produced in \collThree{p}{d}{A} and \coll{A}{A} collisions at RHIC and LHC. These results are computed using a convolved elastic and radiative energy loss model, which receives small system size corrections to both the elastic and radiative energy loss. We observe that suppression in small systems is almost entirely due to elastic energy loss; furthermore, we find that our model is acutely sensitive to the transition between hard thermal loop and vacuum propagators in the elastic energy loss. Finally, we consider the central limit theorem approximation, which is commonly used to model the elastic energy loss distribution as Gaussian.
}
\maketitle
\section{Introduction}
\label{sec:intro}

Over the past decade, the observation of quark-gluon plasma (QGP) signatures in small collision systems has created considerable interest within the heavy-ion physics community. Traditional QGP signatures such as quarkonium suppression \cite{ALICE:2016sdt}, strangeness enhancement \cite{ALICE:2013wgn}, and elliptic flow \cite{CMS:2012qk}---typically associated with large heavy-ion collisions---have also been detected in small systems. These findings raise questions about whether QGP forms in even the smallest collision systems.

One of the most robust probes of QGP formation is jet quenching, often measured in terms of the nuclear modification factor $R_{AB}$ for the collision \coll{A}{B}. This factor quantifies the degree of energy loss experienced by high-$p_T$ partons as they traverse the medium. A measured $R_{AB} \sim 0.2$ for leading pions in central \coll{Au}{Au} collisions at RHIC \cite{PHENIX:2001hpc} has been attributed to significant medium-induced partonic energy loss. By contrast, photons, which do not undergo strong interactions in the medium, exhibit $R_{AB} \sim 1$ \cite{PHENIX:2005yls}.

In small systems, determining $R_{AB}$ is more challenging due to centrality bias, leading to an ambiguous measured suppression pattern. Centrality bias refers to a non-trivial correlation between the hard and soft modes of the QGP, potentially leading to an inaccurate normalization for the $R_{AB}$. This issue is especially pronounced in small systems. LHC data from central \coll{p}{Pb} collisions finds $R_{AB} \simeq 1 \text{--} 1.2$ for pions and $D$ mesons \cite{ALICE:2016yta, ATLAS:2022kqu} at low- to moderate-$p_T$, suggesting no QGP formation. Conversely, PHENIX observes $R_{AB} \simeq 0.75$ in central \coll{d}{Au} collisions, qualitatively consistent with QGP formation \cite{PHENIX:2023dxl}. 

The situation on the theoretical front, however, has its own set of challenges. Theoretical energy loss models generally rely on approximations valid only for large systems. For instance, the Djordjevic-Gyulassy-Levai-Vitev (DGLV) framework for radiative energy loss \cite{Djordjevic:2003zk} assumes a large pathlength $L$, neglecting terms proportional to $e^{-\mu L}$, where $\mu$ is the Debye mass. A \emph{short pathlength correction} (SPL corr.) has been derived that restores these terms \cite{Kolbe:2015rvk}.  Similar large system size assumptions appear in elastic energy loss models, which often assume a Gaussian energy loss distribution based on the central limit theorem, which is inapplicable in small systems \cite{Wicks:2008zz}. In this work, we examine a more realistic Poisson distribution for elastic energy loss \cite{Wicks:2008zz}---which is valid for all system sizes---and compare it to the Gaussian distribution at the level of $R_{AB}$. Furthermore we interrogate a particular uncertainty in the elastic energy loss, which stems from the transition between HTL and vacuum propagators, by computing results with the Braaten and Thoma (BT) elastic energy loss model \cite{Braaten:1991we} as well as results with the HTL elastic energy loss model \cite{Wicks:2008zz}. The BT result utilizes HTL propagators for small momentum transfer and vacuum propagators for large momentum transfer, while the HTL result uses HTL propagators for all momentum transfers.

\section{Results}
\label{sec:results}

We present a model for perturbative QCD energy loss which includes both radiative and elastic contributions. The radiative component is modelled using the DGLV framework, with and without short pathlength corrections (DGLV and DGLV+SPL) \cite{Djordjevic:2003zk, Kolbe:2015rvk}. For the elastic component, we compare results using BT \cite{Braaten:1991we}, Gaussian hard thermal loop (Gauss.\ HTL), and Poisson hard thermal loop (Poiss.\ HTL) models \cite{Wicks:2008zz}. Thermodynamic and geometrical quantities are calculated from IP-Glasma initial conditions \cite{Schenke:2020mbo} coupled with subsequent Bjorken expansion. A detailed description of the model may be found in our other works \cite{Faraday:2023mmx, Faraday:2024gzx}.

\begin{figure}[!t]
	\includegraphics[width=\linewidth]{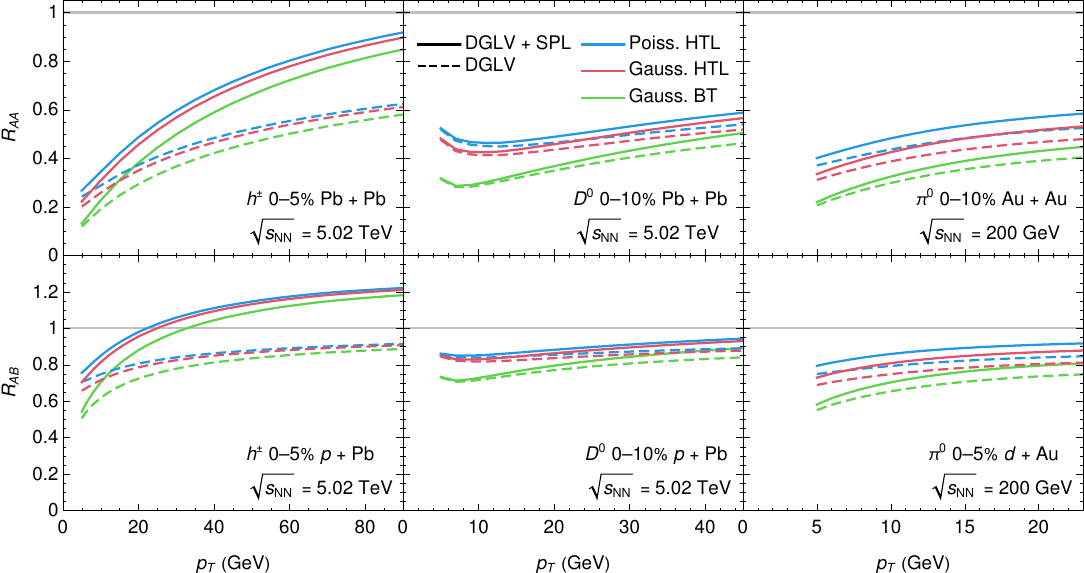}
	\caption{
The top row displays $R_{AA}$ results from our model for charged hadrons produced in $0\text{--}5\%$ centrality \coll{Pb}{Pb} collisions at $\sqrt{s_{NN}} = 5.02 ~\mathrm{TeV}$; $D^0$ mesons produced in $0\text{--}10\%$ centrality \coll{Pb}{Pb} collisions at $\sqrt{s_{NN}} = 5.02 ~\mathrm{TeV}$; and neutral pions produced in $0\text{--}10\%$ centrality \coll{Au}{Au} collisions at $\sqrt{s_{NN}} = 200~\mathrm{GeV}$. The bottom row shows $R_{AB}$ for charged hadrons produced in $0\text{--}5\%$ centrality \coll{p}{Pb} collisions at $\sqrt{s_{NN}} = 5.02 ~\mathrm{TeV}$; $D$ mesons produced in $0\text{--}10\%$ centrality \coll{p}{Pb} collisions at $\sqrt{s_{NN}} = 5.02 ~\mathrm{TeV}$; and neutral pions produced in $0\text{--}5\%$ \coll{d}{Au} collisions at $\sqrt{s_{NN}} = 200 ~\mathrm{GeV}$. Model results are produced by varying the radiative energy loss between the DGLV \cite{Djordjevic:2003zk} and DGLV + SPL models \cite{Kolbe:2015rvk}, and separately the elastic energy loss between the BT \cite{Braaten:1991we}, Gaussian HTL, and Poisson HTL models \cite{Wicks:2008zz}.
	}
	\label{fig:raa_grid}
\end{figure}

In \cref{fig:raa_grid}, we compare $R_{AB}$ results for pions and $D$ mesons produced in central \coll{Pb}{Pb}, \coll{Au}{Au}, \coll{p}{Pb}, and \coll{d}{Au} collisions at RHIC and LHC. Details are found in the caption of \cref{fig:raa_grid}.
There are six model results presented which are calculated by varying the radiative energy loss between the DGLV and the DGLV + SPL models, and separately the elastic energy loss between the BT, Poisson HTL and Gaussian HTL models.

From \cref{fig:raa_grid}, we first compare the $R_{AB}$ calculated with the DGLV and DGLV + SPL radiative energy loss models to understand the impact of the SPL correction. We observe that the SPL correction greatly reduces the suppression for pions in both \coll{Pb}{Pb} and \coll{p}{Pb} collisions; however, it is negligible for $D$ mesons in both \coll{p}{Pb} and \coll{Pb}{Pb} collisions as well as pions in \coll{Au}{Au} and \coll{d}{Au} collisions. 
The large correction for pions at LHC is due primarily to the large proportion of gluons which fragment to pions at LHC compared to RHIC, while $D$ mesons fragment from charm quarks. The SPL correction is significantly larger for gluons compared to quarks, breaking the usual $C_A / C_F = 9 /4$ colour scaling \cite{Kolbe:2015rvk, Faraday:2023mmx}. Furthermore, the SPL correction scales as $\Delta E_{\text{SPL}} \sim E$ while the uncorrected DGLV result scales as $\Delta E_{\text{DGLV}} \sim \ln E$, which explains why the SPL correction grows as a function of $p_T$ \cite{Kolbe:2015rvk}. We also observe that the SPL correction is fractionally larger in small systems compared to larger systems due to the $e^{- \mu L}$ scaling of the SPL correction \cite{Kolbe:2015rvk, Faraday:2023mmx}.

In \cref{fig:raa_grid}, we additionally compare the $R_{AB}$ calculated with the BT and Gauss.\ HTL elastic energy loss models, which is sensitive to the uncertainty in the transition between HTL and vacuum propagators. We observe that this uncertainty leads to an $\mathcal{O}(30\text{--}80)\%$ effect in large systems and an $\mathcal{O}(20\text{--}40)\%$ effect in small systems. Finally, we compare the Poiss.\ HTL and Gauss.\ HTL results to understand the effect of the commonly used central limit theorem approximation. We observe that this effect is $\mathcal{O}(5\text{--}10)\%$ in large systems and $< 2 \%$ in small systems. 
This agreement between Poisson and Gaussian results cannot be attributed to convergence under the central limit theorem, as one would expect the results to align more closely in larger systems than in smaller ones.
In actuality, the agreement is because the $R_{AB}$ in small systems depends mostly on the zeroth and first moments of the elastic energy loss distributions---which are constrained to be the same for Gaussian and Poisson distributions---while in large systems the radiative energy loss dominates over the elastic \cite{Faraday:2024gzx}.

\section{Conclusions}
\label{sec:conclusions}

We have presented suppression results for high-$p_T$ $D$ and $\pi$ mesons produced in \collThree{p}{d}{A} and \coll{A}{A} collisions at RHIC and LHC, from a model which receives small system size corrections to both the radiative and elastic energy loss. We found that an uncertainty in the transition between HTL and vacuum propagators in the elastic energy loss imparts a $\mathcal{O}(40 \text{--} 80)\%$ uncertainty on the $R_{AB}$ for all systems and final states at low- to moderate-$p_T$. Furthermore, we found that the application of the central limit theorem to model the elastic distribution as Gaussian produced a negligible effect at the level of the $R_{AB}$, particularly so in small systems. We attributed this to the fact that for small suppression the $R_{AB}$ depends mostly on the zeroth and first moments of the elastic energy loss, which are constrained to be identical for Gaussian and Poisson distributions, while for larger momenta radiative energy loss dominates.

Future work may involve fitting the strong coupling $\alpha_s$ to data to determine whether observed small and large system suppression can be simultaneously described by a single model. Additionally, one may further analyse the sensitivity to various uncertainties in elastic and radiative energy loss, including the large formation time assumption, which we previously found was not satisfied self-consistently in the DGLV model \cite{Faraday:2023uay, Faraday:2023mmx}. Moreover, placing energy loss models on a more rigorous theoretical foundation could involve carefully treating the kinematics \cite{Clayton:2021uuv}, computing the scales at which the coupling runs at finite temperature and system size, and understanding the validity of various approximations in the calculations \cite{Horowitz:2009eb}. Lastly, investigating QGP formation in small systems may also include examining the impact of small system size on the coupling \cite{Horowitz:2022rpp}, thermodynamics \cite{Mogliacci:2018oea}, the equation of state \cite{Horowitz:2021dmr}, and jet substructure observables \cite{Kolbe:2023rsq}.

\section*{Acknowledgements}
CF and WAH thank the South African National Research Foundation and SA-CERN Collaboration for financial support.

% \bibliography{manual.bib, 315_Parallel_HeavyFlavourEnergyLossInSmallAndLargeSystems_ColeridgeFaraday_v2024-Sep-22.bib}

\end{document}